\documentclass[prl]{revtex4}%
\usepackage{amsmath}
\usepackage{amsfonts}
\usepackage{amssymb}
\usepackage{graphicx}%
\begin{document}

\title{Pseudo-Rabi oscillations in superconducting flux qubits in the classical regime}

\author{A.N. Omelyanchouk$^1$, S.N. Shevchenko$^1$, A.M. Zagoskin$^{2,3}$, E. Il'ichev$^4$, Franco Nori$^{2,5}$}
\affiliation{$^1$B.Verkin Instiute for Low Temperature Physics and Engineering, 47 Lenin Ave., 61103, Kharkov, Ukraine,\\ $^2$Digital Materials Laboratory, Frontier Research System,
RIKEN, Wako-shi, Saitama 351-0198, Japan,\\ $^3$Department of Physics and Astronomy,
The University of British Columbia,
Vancouver, B.C., V6T 1Z1 Canada,\\ $^4$Institute of Photonic Technology (IPHT-Jena),
P.O. Box 100239, D-07702 Jena, Germany,\\ $^5$Center for Theoretical Physics, Physics Department,
Center for the Study of Complex Systems,
The University of Michigan, Ann Arbor, MI 48109-1040, USA}

\begin{abstract}
Nonlinear effects in mesoscopic devices can have both quantum and classical origins.
We show that a three-Josephson-junction (3JJ) flux qubit in the {\em classical} regime can produce low-frequency oscillations in the presence of an external field in resonance with the (high-frequency) harmonic mode of the system, $\omega$.  Like in the case of {\em quantum} Rabi oscillations, the frequency of these pseudo-Rabi oscillations  is much smaller than $\omega$ and scales approximately linearly with the amplitude of the external field.
This classical effect can be reliably distinguished from its quantum counterpart because it can be produced by the external perturbation not only at the resonance frequency $\omega$ and its subharmonics ($\omega/n$), but also at its overtones, $n\omega$.

\end{abstract}

\maketitle

The  many advances in  understanding the quantum behaviour of meso- and macroscopic systems over the last decade are due, to a large extent, to the investigation of superconducting qubits in the context of quantum computing~\cite{You2005a,Wendin2006}. These led to the demonstration  of their  quantum behavior, at the level of one to four qubits~\cite{Nakamura1999,Wal2000a,Izmalkov2004a,Izmalkov2006,Grajcar2006}.

Since flux qubits can exhibit quantum superpositions of states differing by a macroscopic number of  single-electron states, and the relevant observables can be easily accessed experimentally, these devices provide unique opportunities to investigate the quantum-classical frontier.
Further scaling up of superconducting qubit networks is thus motivated by the needs of  quantum information processing in solid state as well as by the fundamental interest     in  probing the limits of the applicability of quantum mechanics.

Once measurements are taken into account, the quantum behaviour becomes essentially nonlinear.  However, already at the classical level, nonlinearities are unavoidable  in  superconducting qubits, due to the nonlinear behaviour  of  Josephson
junctions.
It was  recently pointed out that in a phase qubit, which is a biased single Josephson junction, the classical nonlinearity can produce  effects with  characteristics similar to  Rabi oscillations~\cite{Gronbech-Jensen2005} and Ramsey fringes ~\cite{Marchese2006}, which are often considered as   signatures of quantum behaviour in two-level systems

The results of ~\cite{Gronbech-Jensen2005,Marchese2006} do not undermine the common understanding that superconducting qubits demonstrate quantum behaviour, since the latter was established by a set of independent methods~\cite{Wendin2006}. Rather it attracts our attention to the interesting possibility of coexistence of similar nonlinear classical and quantum  effects in the same device. How to distinguish classical versus quantum behavior in these qubits is an important question and the focus of this work.

 Rabi oscillations (e.g., Ref.~\cite{Meystre1991}, p.~89) are coherent quantum transitions in a two-level system  driven by an external ac field of amplitude A and with the characteristic frequency
 \begin{equation}
    \Omega = \sqrt{A^2 + (\delta\omega)^2},
\end{equation}
 where  $\delta\omega = |\omega - \omega_0|$ is its detuning from the interlevel distance, $\omega_0$. In resonance, $\Omega = A$. This linear dependence on the field amplitude, and $\Omega$ being much less than other characteristic frequencies in the system, help identify Rabi oscillations. Indeed, they were observed in all types of superconducting qubits~\cite{Wendin2006}, and their appearance is sometimes used as  decisive evidence in favour of quantum behaviour. Multiphoton Rabi oscillations, at $\omega_0 = n\omega$, were also observed ~\cite{Saito2005}.

In this paper we investigate a three-Josephson-junction (3JJ) flux qubit in the classical regime. With {\em two} independent variables instead of one, this is a richer system, than the phase qubit of ~\cite{Gronbech-Jensen2005}. We find that the resonant high-frequency driving $\omega$ produces  low-frequency oscillations of the magnetic flux which are very similar to  Rabi oscillations.  We also show that a {\em qualitative} difference exists between these two effects, which allows to reliably distinguish them in experiments.

Following  Ref.~\cite{Mooij1999}, we consider a 3JJ flux qubit in the limit of negligible self-inductance $L\rightarrow0$. The magnetic flux through the loop then equals the external applied flux $\Phi
_{e},$ and we introduce the reduced flux $\varphi_{e}=2\pi\Phi_{e}/\Phi_{0}$. Here $\Phi_0 = h/2e$ is the flux quantum. Bistability is achieved due to the presence of three junctions in the loop. Due to the single-valuedness of the superconducting wave function, one of the phase shifts across the Josephson junctions is eliminated through
$\varphi_{1}+\varphi_{2}+\varphi_{3}=\varphi_{e},
$
leaving two independent variables,
$
\theta = (\varphi_1+\varphi_2)/2$ and $\chi = (\varphi_1-\varphi_2)/2.
$

\begin{figure}
    \centering
        \includegraphics{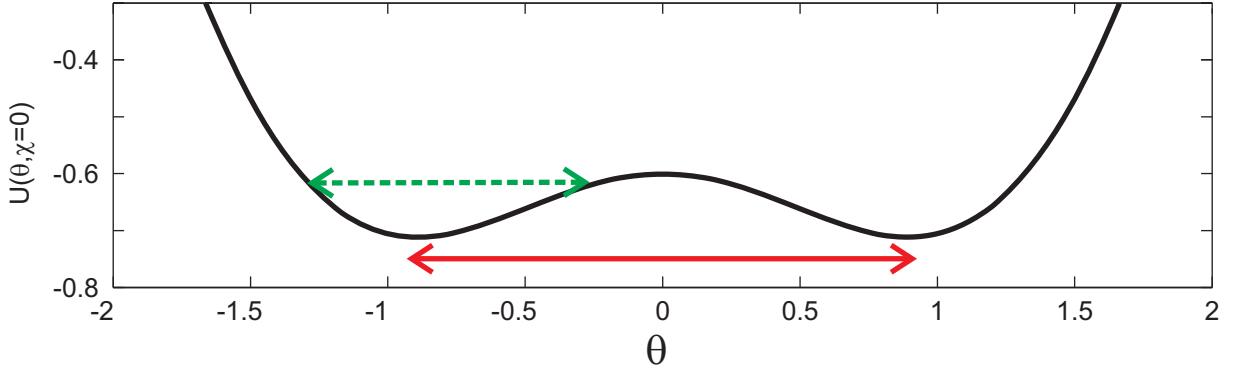}
    \caption{(Color online) The potential profile of Eq.~(\ref{eq_potential}) with $\alpha=0.8, \:\varphi_e^d = \pi$. The arrows indicate quantum (solid) and classical (dotted) oscillations.}
    \label{fig1}
\end{figure}

In the classical regime, the phase dynamics of the $i$th Josephson junction ($i=1,2,3$) can be described by the RSJJ model~\cite{Barone1982}, in which the current is given by

\begin{equation}
I=\frac{\hbar C_{i}}{2e}\frac{d^{2}}{dt^{2}}\varphi_{i}+\frac{\hbar}{2eR_{i}%
}\frac{d}{dt}\varphi_{i}+I_{ci}\sin\varphi_{i}. \label{eq_I}
\end{equation}
Here $C_i$ is the junction capacitance, $R_i$ its normal resistance, and $I_{ci}$ its critical current. We now neglect the effect of thermal fluctuations, which are not crucial here. Following the common choice of parameters, we set $
C_{1}=C_{2}=C,\: I_{c1}=I_{c2}=I_{c},\:R_{1}=R_{2}=R;\:\:$
$ C_{3}=\alpha C,\:I_{c3}=\alpha I_{c},\:R_{3}=R/\alpha;\:(0.5 \leq \alpha \leq 1).
$
 Introducing the dimensionless units, $ \omega_{0}t=\tau,\:\gamma=
\hbar\omega_{0}/2eRI_{c} = \omega_{0}/\omega_{R},
$
where $\omega_{0}=\sqrt{(2eI_{c})/(\hbar C)},\:$ and $\omega_R = (2eRI_{c})/\hbar$,
we can write the equations of motion for the variables $\theta,\:\chi$:
\begin{eqnarray}
\frac{d^{2}}{d\tau^{2}}\chi+\gamma\frac{d}{d\tau}\chi   = -\cos\theta\sin
\chi,\nonumber\\
(1+2\alpha)\frac{d^{2}}{d\tau^{2}}\theta+\gamma(1+2\alpha)\frac{d}{d\tau
}\theta   = \label{eq_motion}\\
  -\sin\theta\cos\chi+\alpha\sin(\varphi_{e}-2\theta)+\alpha\frac{d^{2}%
}{d\tau^{2}}\varphi_{e}+\alpha\gamma\frac{d}{d\tau}\varphi_{e}\:.%
\nonumber
\end{eqnarray}
We consider the external flux
\begin{equation}
\varphi_{e}(t)\: \equiv \: 2\pi(\Phi_e/\Phi_0) \: = \: \varphi_{e}^{d}+\varphi_{e}^{a}\sin(\omega\tau). \label{eq_external}
\end{equation}
The energy of the system is thus
\begin{equation}
H  =E_J \left[\frac{1}{2}\left(\frac{d}{d\tau}\chi\right)^{2}+\frac{1}{2}(1+2\alpha)\left(\frac
{d}{d\tau}\theta\right)^{2}   -\cos\theta\cos\chi-\frac{1}{2}\alpha\cos(\varphi_{e}-2\theta)\right],
\end{equation}
where $E_{J}=\hbar I_{c}/2e$ is the Josephson energy. The canonical momenta are
$p_{\chi}=E_J d\chi/d\tau$, $p_{\theta}=E_J(1+2\alpha) d\theta/d\tau.$ The effective potential is given by (Fig.~\ref{fig1})

\begin{equation}
U(\theta,\chi) = -\cos\theta \cos\chi - \alpha \cos(\varphi_e - 2\theta)/2.
\label{eq_potential}
\end{equation}

If the dc static bias is $\varphi_e^d = \pi$, the system has degenerate minima at $\theta_{0}=\pm \arccos\left[(2\alpha)^{-1}\right],\chi_{0}=0.$ The ``plasma'' frequencies of small oscillations around them are (in units of $\omega_0$) $
\omega_{\theta}=\sqrt{1-(2\alpha)^{-1}}, \:\omega_{\chi}=(2\alpha)^{-1/2}$.

\begin{figure}
    \centering
        \includegraphics{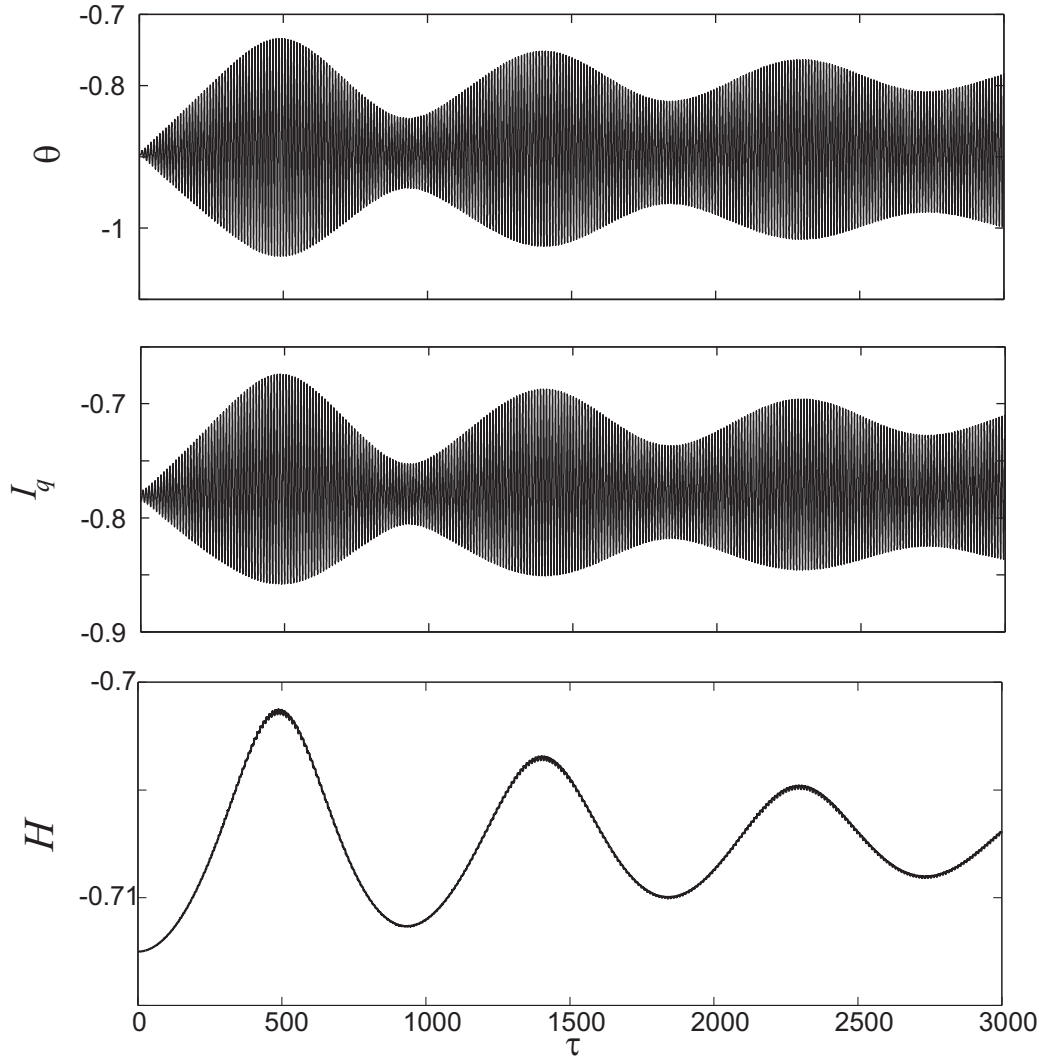}
            \caption{(a) Driven oscillations around a minimum of the potential profile of Fig. \ref{fig1} as a function of time. The driving amplitude is $\varphi_e^a = 0.01$, driving frequency $\omega = 0.612$, and the decay rate $\gamma = 10^{-3}$. Low-frequency classical beat oscillations are clearly seen. (b)  Low-frequency oscillations of the persistent current in the  3JJ loop. (c) Same for the energy of the system.
    }
    \label{fig2}
\end{figure}

\begin{figure}
    \centering
        \includegraphics{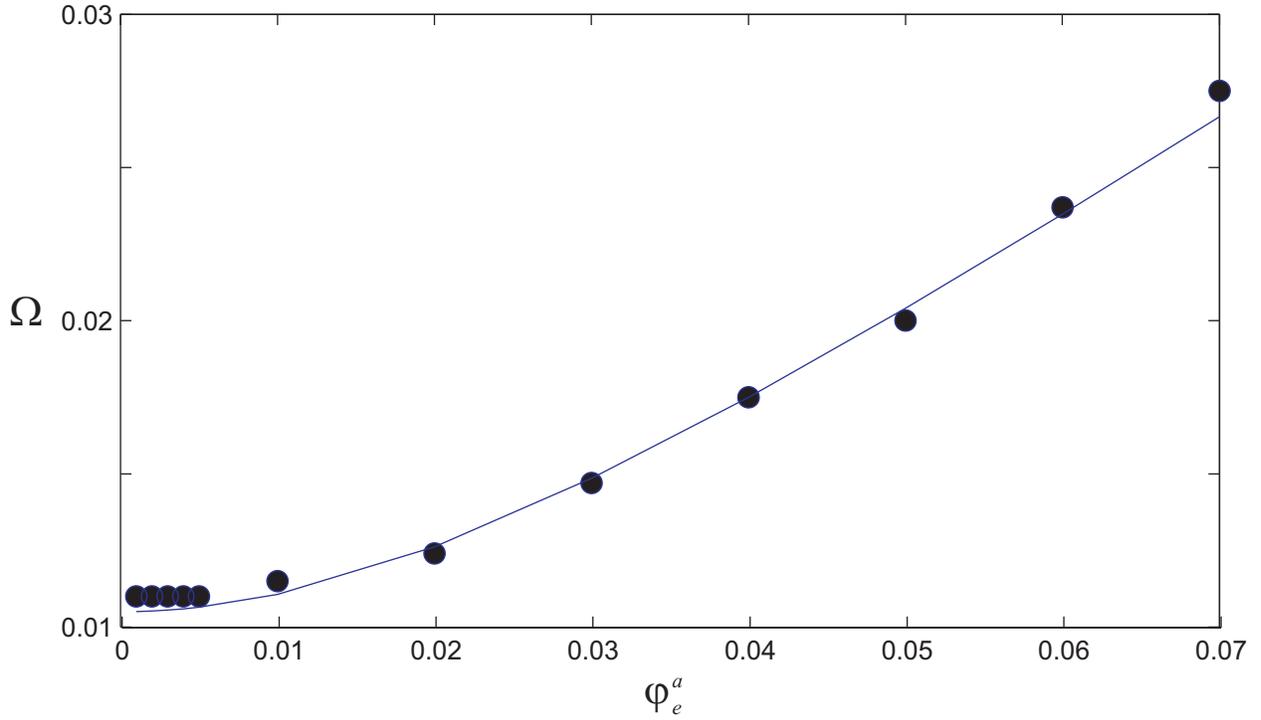}
    \caption{The dependence of the pseudo-Rabi frequency on the driving amplitude $\varphi_e^a$ for $\omega = 0.6,\: \gamma = 10^{-3}$. The solid line, $\Omega = 0.35  \left[\left(\varphi_e^a\right)^2 + \left(\omega-0.63\right)^2\right]^{1/2}\!\!,$  is the best fit to the calculated data.}
    \label{fig3}
\end{figure}

In the presence of the external field (\ref{eq_external}) the system will undergo forced oscillations around one of the equlibria. For   $\alpha=0.8$~\cite{Mooij1999}, which is close to the parameters of the actual devices~\cite{Chiorescu2003,Ilichev2003,Grajcar2006}, the values of the dimensionless frequencies become $\omega_{\theta} \approx 0.612,$ and $\omega_{\chi} \approx 0.791.$ Solving the equations of motion (\ref{eq_motion}) numerically, we  see the appearance of slow oscillations of the amplitude and energy superimposed on the fast forced oscillations (Fig.~\ref{fig2}), similar to the classical oscillations in a phase qubit (Fig.~2 in ~\cite{Gronbech-Jensen2005}). The dependence of the  frequency of these oscillations on the driving amplitude shows an almost linear behaviour
(Fig.~\ref{fig3}), which justifies the ``Pseudo-Rabi'' moniker.

\begin{figure}
    \centering
        \includegraphics{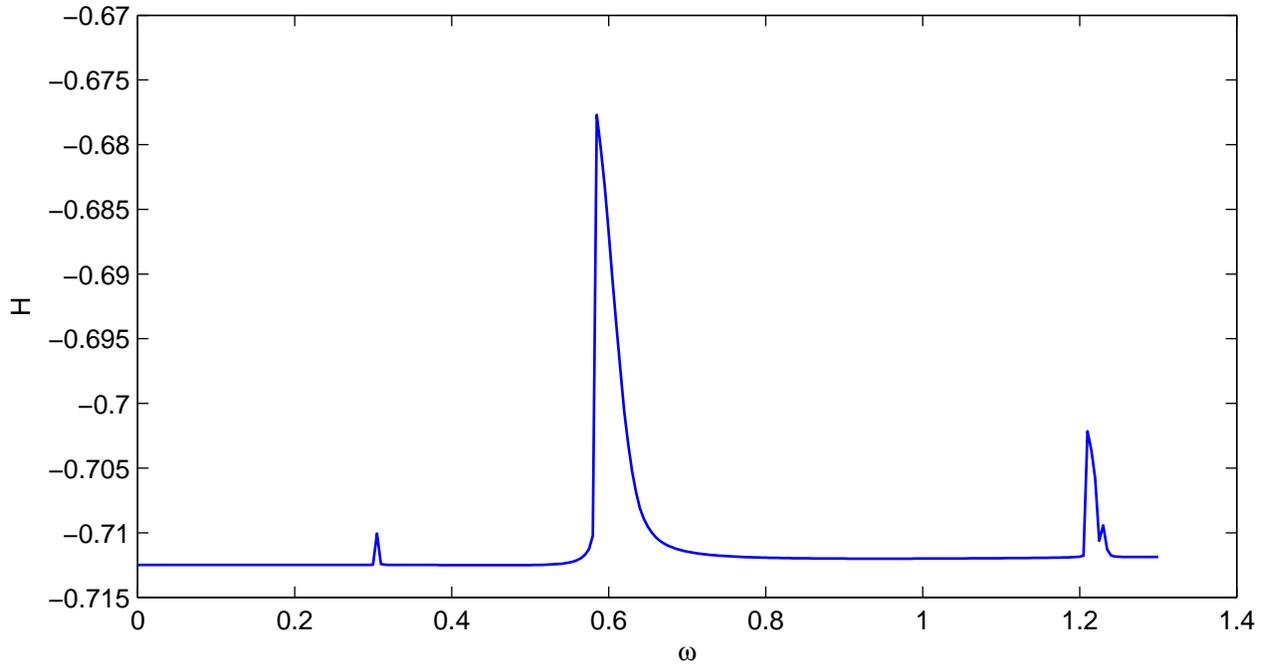}
    \caption{The average energy $H$ of the system as a function of the driving frequency $\omega$. The main peak ($\omega_0 \approx 0.6$) corresponds to the resonance. The left peak at $\omega_0/2$ is the nonlinear effect of the excitation by a subharmonic, similar to a multiphoton process in the quantum case. The right peak at $2\omega_0$ is the first overtone and it has {\em no} quantum counterpart. Here $\varphi_e^d= \pi; \: \varphi_e^a = 0.05;\:\gamma = 10^{-3}.$}
    \label{fig4}
\end{figure}

A quantitative difference between this effect and true Rabi oscillations is in the different scale of the resonance frequency. To induce  Rabi oscillations between the lowest quantum levels in the potential (\ref{eq_potential}), one must apply a signal in resonance with their tunneling splitting, which is exponentially smaller than $\omega_0$. Still, this is not a very reliable signature of the effect, since the classical effect can also be excited by subharmonics, $\sim \omega_0/n$, as we can see in Fig.~\ref{fig4}.

The key observable difference between the classical and quantum cases, which would allow to reliably distinguish between them, is that the classical effect can also be produced by driving the system at the {\em overtones} of the resonance signal, $\sim n\omega_0$ (Fig.~\ref{fig4}). This effect can be detected using  a standard technique for   RF SQUIDs~\cite{Golubov2004}.
The current circulating in the qubit circuit produces a magnetic moment, which is measured by the inductively coupled high-quality tank circuit. For the tank voltage $V_T$ we have
\begin{equation}
\frac{d^2V_T}{dt^2} + \frac{1}{\tau_T}\frac{dV_T}{dt} + \omega_T^2 V_T = \omega_T^2 M \frac{dI_q}{dt}, \label{eq_VT}
\end{equation}
where $\tau_T = R_TC_T$ is the RC-constant of the tank, $\omega_T = (L_TC_T)^{-1/2}$ its resonant frequency, $M$ the mutual inductance between the tank and the qubit, and $I_q(t)$ the current circulating in the qubit. The persistent current in the 3JJ loop can be determined directly from (\ref{eq_I}).
Its behaviour in the presence of an external RF field is shown in Fig.~\ref{fig2}c. Note that the sign of the current does {\it not} change, which is due to the fact that the oscillations take place {\it inside} one potential well (solid arrow in Fig.~\ref{fig1}), and not {\em between} two separate nearby potential minima like in the quantum case. (Alternatively, this would also allow to distinguish between the classical and quantum effects by measuring the magnetization with a DC SQUID.)

There can be two strategies in detecting the slow oscillations, as long as the measurement time is smaller than their decay time. First, one can directly measure the time-dependent voltage in the tank circuit, which from (\ref{eq_VT}) is
\begin{equation}
V_T(\omega) = \frac{i \omega \omega_T^2 M}{\omega^2 - \omega_T^2 + i\omega/\tau_T} I_q(\omega). \label{eq_induced}
\end{equation}
Alternatively, one can measure the spectral density of the signal in the tank,
\begin{equation}
\langle V_T^2\rangle_{\omega} = \frac{\omega^2\omega_T^4 M^2}{(\omega^2-\omega_T^2)^2 + \omega^2/\tau_T^2} \langle I_q^2\rangle_{\omega}.
\end{equation}
Choosing  the tank frequency $\omega_T$ close to the classical low ``Rabi'' frequency $\Omega$,
in either case we use the tank as a filter, which removes the
interference from the large, high-frequency driving field. Note
that in the first case we have an additional discriminant of the
quantum versus classical behaviour of the system. In the classical
case, the time-dependent regular oscillations of the tank voltage
or current due to the oscillations in the qubit,
(\ref{eq_induced}), can be measured directly. In the quantum case,
this is impossible, due to the uncertainty principle, and the
oscillations are only seen in the statistics of the measurements.
Remarkably, the effect on the correlators can be directly observed
in both quantum and classical cases (in the quantum case such
measurements are not limited by the decay time
~\cite{Ilichev2003}).

In conclusion, we predict that  a 3JJ flux qubit driven by a resonant external field will exhibit classical low-frequency oscillations superficially-similar to the quantum Rabi oscillations in a driven two-level system. Both effects can coexist in the same region of parameters. A qualitative difference between the two (allowing to reliably distinguish between them) is that the classical effect can be driven also at the overtones of the resonant frequency.

We acknowledge partial support from the NSA, LPS, ARO, NSF grant
No. EIA-0130383, JSPS-RFBR 06-02-91200, MEXT
Grant-in-Aid No.~18740224, JSPS-CTC program, the NSERC Discovery Grants Program (Canada), RSFQubit and EuroSQIP. SNS acknowledges the financial support by INTAS under the YS Fellowship Grant. ANO and SNS are grateful to DML FRS RIKEN for their hospitality.


\end{document}